\documentclass[conference]{IEEEtran}
\IEEEoverridecommandlockouts
\usepackage{cite}
\usepackage{amsmath,amssymb,amsfonts}
\usepackage{algorithmic}
\usepackage{graphicx}
\usepackage{textcomp}
\usepackage{xcolor}
\usepackage{url}

\usepackage{multirow}
\usepackage{makecell}
\usepackage{siunitx}
\usepackage{blindtext}
\usepackage{multicol}
\usepackage{algorithm}
\usepackage{subfig}
\usepackage{booktabs}
\usepackage{subfloat}
\usepackage{caption}
\usepackage{subfiles}
\usepackage{subcaption}
\usepackage[table]{xcolor}

\def\BibTeX{{\rm B\kern-.05em{\sc i\kern-.025em b}\kern-.08em
    T\kern-.1667em\lower.7ex\hbox{E}\kern-.125emX}}
\begin{document}

\title{MLLM-DataEngine: Closing the Loop of Multimodal Instruction Tuning Data Generation
\thanks{
* Equal contribution.
}
\thanks{
$\dag$ Corresponding author.
}
\thanks{
This project was supported by National Key R\&D Program of China (NO.2022ZD0160102) and Shanghai Artificial Intelligence Laboratory.
}
}

\author{
\IEEEauthorblockN{Zhiyuan Zhao\IEEEauthorrefmark{1}, Bin Wang\IEEEauthorrefmark{1}, Linke Ouyang\IEEEauthorrefmark{1}, Yiqi Lin, Pan Zhang, Xiaoyi Dong, Jiaqi Wang, Conghui He\IEEEauthorrefmark{2}}
\IEEEauthorblockA{\textit{Shanghai AI Laboratory} \\
Shanghai, China \\
\{zhaozhiyuan,~wangbin,~ouyanglinke,~dongxiaoyi,~heconghui\}@pjlab.org.cn, wjqdev@gmail.com, linyq29@gmail.com}
}

\maketitle

\begin{abstract}
In this paper, we propose MLLM-DataEngine, a novel closed-loop system that bridges data generation, model training, and evaluation. Within each loop iteration, the MLLM-DataEngine first analyzes the weakness of the model based on the evaluation results, then generates a proper incremental dataset for the next training iteration, and enhances the model capability iteratively. Compared with previous instruction fine-tuning dataset collection methods which are separate from the benchmarking, MLLM-DataEngine shows better targeting and can improve MLLMs's capabilities more effectively. Firstly, we propose an Adaptive Bad-case Sampling module, which can effectively analyze model weakness based on the benchmarking results and adjust the generation of incremental datasets flexibly. Secondly, in order to ensure high-quality data for specific capability types, the most representative in-context examples and abundant information are provided to GPT-4, which helps GPT-4 fully comprehend the model's weakness and further guarantees high-quality generated data. Through extensive experiments, we find MLLM-DataEngine could boost the MLLMs capability in a targeted and automatic manner without human participants. We hope MLLM-DataEngine could be a general solution for the following MLLMs data curation. Code, data, and model are available at \url{https://github.com/opendatalab/MLLM-DataEngine}.
\end{abstract}


\begin{figure*}[h]
    \centering
    \subfloat[ChatCaptioner]{\includegraphics[scale=0.33]{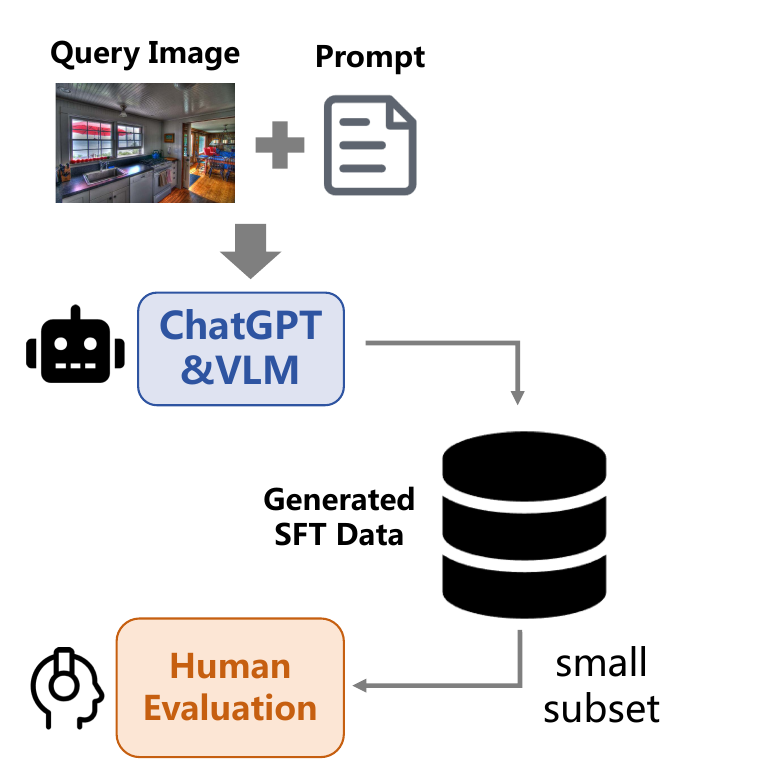}}
    \subfloat[IdealGPT]{\includegraphics[scale=0.33]{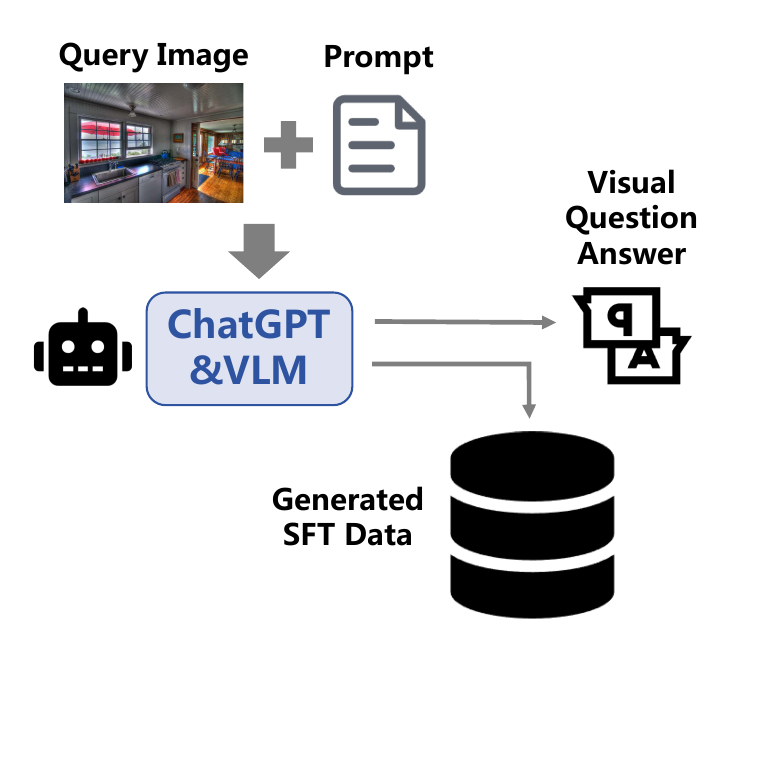}}
    \subfloat[LRV]{\includegraphics[scale=0.33]{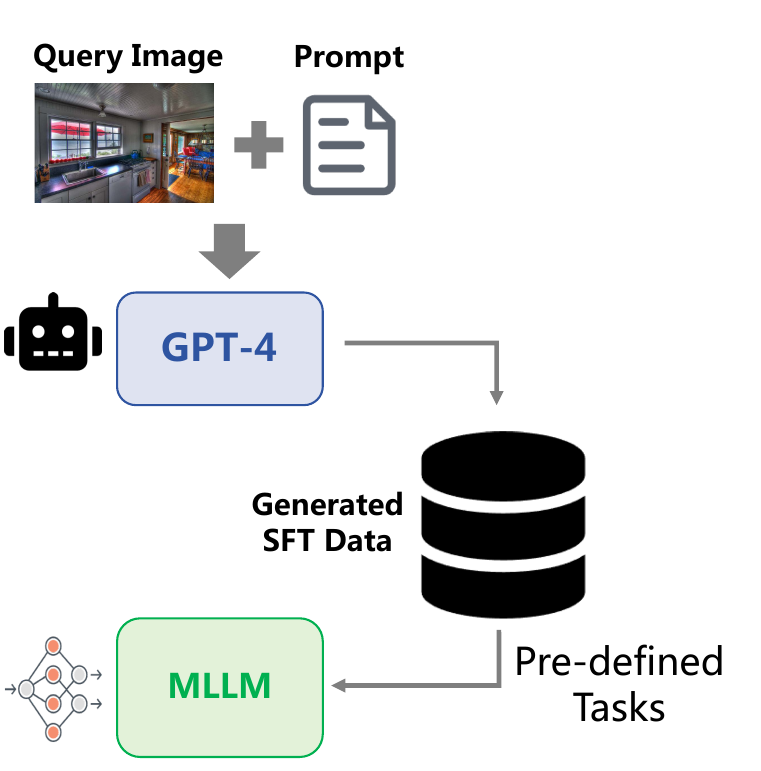}}
    \subfloat[Our Method]{\includegraphics[scale=0.33]{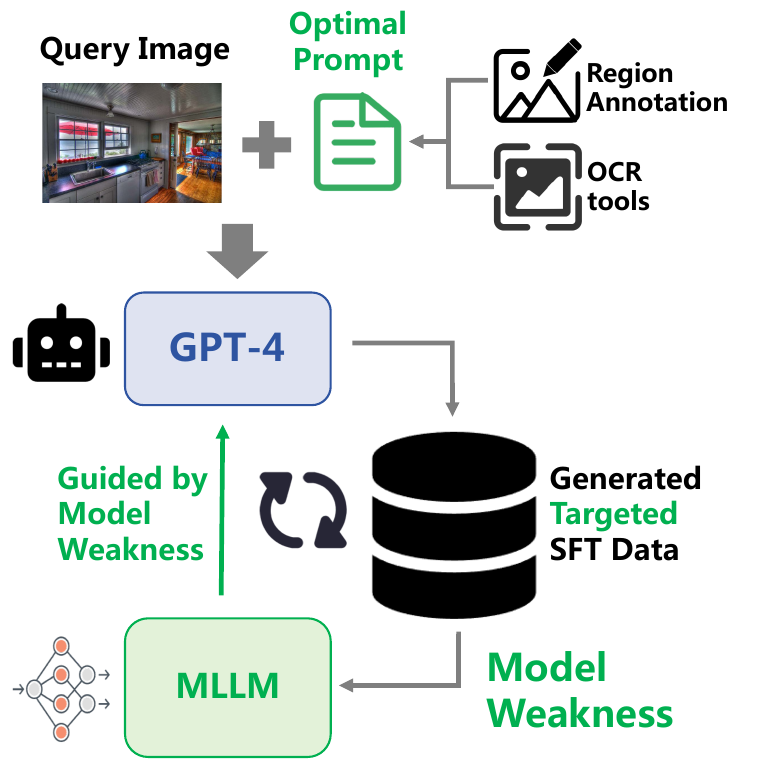}}
    \caption{Comparison of existing methods and our proposed MLLM-DataEngine. Existing instruct tuning data generation methods are separated from model evaluation and feedback. In contrast, our proposed MLLM-DataEngine is a closed-loop of generation-training-evaluation-generation which leads to targeted and effective model improvement.}
    \label{fig:fig1}
\end{figure*}

\section{Introduction}
\label{sec:intro}

The field of Multimodal Large Language Models (MLLMs) has greatly advanced recently~\cite{DBLP:journals/corr/abs-2304-10592, DBLP:journals/corr/abs-2304-08485, DBLP:journals/corr/abs-2305-06500, DBLP:journals/corr/abs-2306-14824, DBLP:conf/nips/AlayracDLMBHLMM22, DBLP:journals/corr/abs-2401-16420, DBLP:journals/corr/abs-2310-03744}.
To equip these models with vision-language understanding capabilities, a two-stage fine-tuning process becomes common practice~\cite{DBLP:journals/corr/abs-2305-06500,DBLP:journals/corr/abs-2304-08485,DBLP:journals/corr/abs-2304-10592}. In the first stage, the image-text feature alignment is inadequate to ensure robust multi-modal understanding capabilities. More importantly, the second stage utilizes high-quality annotated data for instruction fine-tuning, which is pivotal in guaranteeing exceptional model performance. 

In the pursuit of high-quality multi-modal instruction tuning data, recent studies have begun to explore high-quality data generation. For example, several efforts \cite{DBLP:journals/corr/abs-2305-06500, DBLP:journals/corr/abs-2306-05425} have been undertaken to construct data from public datasets hand-craftly. 
In contrast, some recent advancements like ChatCaptioner \cite{DBLP:journals/corr/abs-2303-06594} and IdeaGPT \cite{DBLP:journals/corr/abs-2305-14985} using Large Language Models (LLMs) like ChatGPT \cite{DBLP:journals/corr/abs-2303-08774} and a Vision-Language Model (VLM) for creating caption data.
For example, LLaVA \cite{DBLP:journals/corr/abs-2304-08485} harnesses GPT-4 \cite{DBLP:journals/corr/abs-2303-08774}, a superior text model, with image annotation for multimodal data generation.
Similarly, LRV~\cite{liu2023mitigating} pre-defines over 20 tasks and combines them with image annotation, enabling GPT-4 to generate high-quality instruction fine-tuning data. 

Despite their great efforts, current methods remain isolated from model evaluation and feedback, which severely hampers the opportunity for model improvement and fails to effectively mitigate model shortcomings. We argue that to improve model capabilities, the construction of high-quality instruction tuning data was supposed to be closely integrated with proficiency benchmarking. Unfortunately, combining benchmarking with data generation is still a remaining challenge. The current benchmarks for MLLMs such as SEED-Bench~\cite{DBLP:journals/corr/abs-2307-16125} and MMBenchmark~\cite{DBLP:journals/corr/abs-2307-06281} provide comprehensive assessment of model capabilities and can point out the model weakness~\cite{DBLP:journals/corr/abs-2304-02643}, while it is non-trivial to use it as guidance for model improvement, especially when the weakness includes several different aspects. A straightforward solution is to annotate or collect new data by humans, while its cost is quite large, especially when the models and benchmarks are updated iteratively.

To solve this problem, we propose \textbf{MLLM-DataEngine} to bridge instruct tuning data generation, model training, and evaluation.
As shown in Fig.~\ref{fig:fig1}, different from previous approaches, our method introduces a closed-loop cycle of \textbf{Evaluation-Guidance-Generation-Evaluation} where model weaknesses from the evaluation phase are harnessed to guide the data generation process.
Updating MLLMs in the loop can more effectively improve model capabilities and mitigate model weaknesses.
Specifically: \textbf{(1) Evaluation:} To comprehensively identify model weakness, we collect bad cases of each fine-grained capability through wide-coverage evaluation and build a bad case pool in each iteration. \textbf{(2) Guidance: }As an essential connection between model evaluation and data generation, \textbf{A}daptive \textbf{B}ad-case \textbf{S}ampling module~(ABS) is proposed to select proper question types and in-context examples from the bad case pool according to model weakness, which helps guide further data generation. \textbf{(3) Generation} We utilize GPT-4 to generate diverse and accurate instruct-tuning data for each fine-grained capability. To make GPT-4 fully comprehend the weakness of the model, we feed the most representative in-context examples to GPT-4. Meanwhile, rich and sufficient information is provided to GPT-4 to ensure the correctness of the generated data. Our contributions are as follows:
\begin{itemize}
  \item We present MLLM-DataEngine, a multimodal engine that fosters a closed loop for data generation, model training, and evaluation, thus facilitating iterative improvement of model capabilities.

  \item Different from current methods, MLLM-DataEngine innovatively guides data generation using model feedback through a carefully designed process, which helps bridge the gap between model improvement and evaluation.
  
  \item We perform extensive experiments across various evaluation benchmarks. Results confirm MLLM-DataEngine's ability to iteratively enhance model performance and compensate for model deficiencies.
\end{itemize}

\section{Our Approach}

\begin{figure*}[t]
    \centering
    \includegraphics[width=0.96\textwidth]{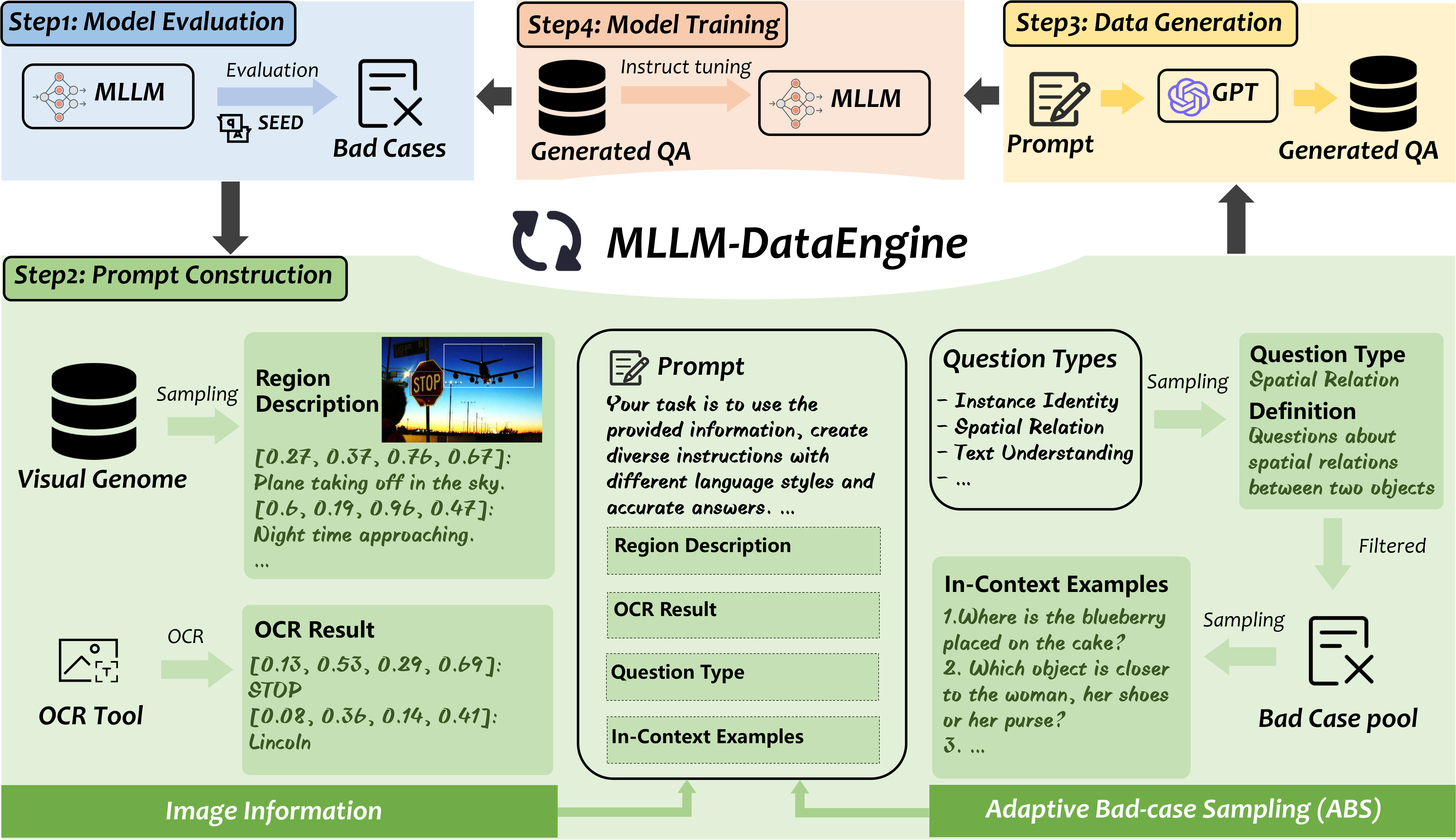}
    \caption{Illustration of proposed MLLM-DataEngine. The whole process is divided into 4 steps. (1) \textbf{Model Evaluation.} We first test the base model on the public benchmark to get the bad cases and build the bad case pool. (2) \textbf{Prompt Construction.} After bad cases are obtained, Adaptive Bad-case Sampling (ABS) is proposed to select the proper question type and the most representative in-context examples. Meanwhile, rich image information is provided. (3) \textbf{Data Generation.} Constructed prompts are fed to GPT-4 to generate data. (4) \textbf{Model Training.} The model is fine-tuned on the latest generated data and loops back to the beginning of the data engine. }
    \label{fig:fig2}
\end{figure*}

The framework of MLLM-DataEngine, showcased in this paper, utilizes feedback from the model evaluation to guide the data generation and harnesses generated data to encounter the weakness of the model, which establishes a cyclical process for iterative enhancement between the model and data. As Fig.~\ref{fig:fig2} illustrates, each iteration involves four steps:

\begin{itemize}
    \item \textbf{Model Evaluation}. Firstly, The model's capabilities are systematically evaluated across various dimensions. Then its bad cases are collected from identified weaknesses.

    \item \textbf{Prompt Construction}. 
    Secondly, proper prompts targeted at model weakness are constructed through Adaptive Bad-case Sampling~(ABS) for GPT-4 data generation.

    \item \textbf{Data Generation}. Thirdly, the prompt constructed in the previous step is fed into GPT-4 for data generation. Meanwhile, all generated data are carefully filtered and processed to ensure high-quality instruction fine-tuning.

    \item \textbf{Model fine-tuning}. Finally, the model is fine-tuned with newly generated data, and MLLM-DataEngine loops back to the first Model Evaluation step for new model performance evaluation and weakness identification. 

\end{itemize}

\subsection{Model Evaluation} \label{Sec: Model Evaluation}

In the first step, we evaluate the model's performance to identify its weaknesses, and we use bad cases (questions that the model answers incorrectly) as feedback to guide further data generation. In our proposed method, we utilize the open-source, Supervised, Fine-Tuned~(SFT) Multimodal Large Language Models (MLLMs) as the initial model. 

As for the evaluation benchmark, we choose SEED-Bench \cite{DBLP:journals/corr/abs-2307-16125} as the evaluation benchmark, given that traditional single-task evaluation methods, such as VQA and Caption, cannot comprehensively and accurately assess the capabilities of MLLMs. 
SEED-Bench is a high-quality and generative evaluation system for multimodal large language models, which involves 19K multiple-choice questions with accurate human annotations. 
Moreover, SEED-Bench evaluation spans over 9 evaluation dimensions for image understanding, including multiple complex vision-language recognition and reasoning tasks~(such as Instance Recognition, Text Recognition, Visual Reasoning, \textit{et al.}), which ensures comprehensive capability evaluation and enhancement in our proposed MLLM-DataEngine.

After the model's performance is comprehensively evaluated, in order to provide targeted guidance for the further data generation process, we collect those bad cases that are answered by MLLMs incorrectly and construct \textbf{bad case pool}. 
The bad case pool contains questions the model answered incorrectly in each type of SEED-Bench evaluation dimension and can reflect the model's shortcomings and defects in specific capability dimensions. During the bad case pool construction, we find that a significant portion of the examples in the bad case pool is similar to each other (such as those about color and \textit{et al.}). 
To make the examples in the bad case pool more representative and diverse, we adopt a simple yet effective duplicate-reduction process to iteratively remove duplicate questions and only keep the most representative examples (see Algorithm~\ref{functionRD}).

\begin{algorithm}[t] 
\caption{Reduce Duplicates in the Bad Case Pool}
\textbf{Input}: $P$ - Original set of bad cases.\\
\textbf{Output}: $Q$ - Reduced bad cases.
\begin{algorithmic}[1]
\STATE Initialize $Q \leftarrow \emptyset$
\WHILE{$P \neq \emptyset$} 
    \STATE $P_{s} = \text{RandomSelect}(P)$  
    \STATE $Q \leftarrow Q \cup P_{s}$  
    \FOR{$P_{i}$ in $P$} 
        \IF{$\text{spacy}(P_{i}, P_{s}) > 0.9$} 
            \STATE $P \leftarrow P.\text{remove}(P_{i})$  
        \ENDIF
    \ENDFOR
\ENDWHILE
\STATE \textbf{return} $Q$
\end{algorithmic} 
\label{functionRD}
\end{algorithm} 

\begin{figure*}[t]
    \centering
    \includegraphics[width=\textwidth]{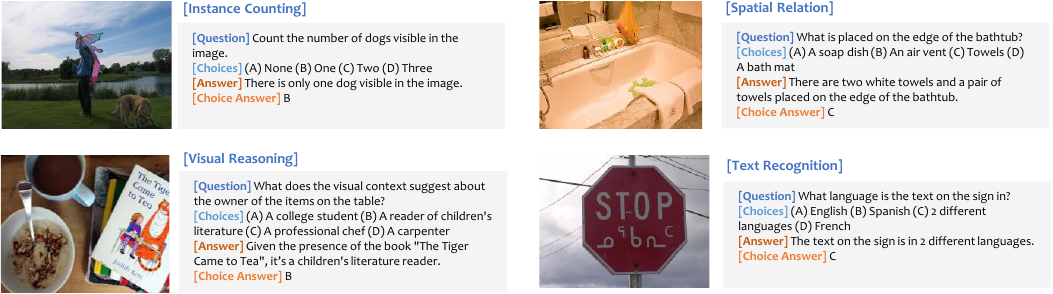}
    \caption{Example of generated data for four out of nine fine-grained abilities.}
    \label{fig:ability}
\end{figure*}

\begin{table*}[t]
    \caption{Experiments on LLaVA-1.5. The None~(baseline) row indicates the baseline model results without incremental dataset in instruction fine-tuning.}
    \centering
    \resizebox{1.0\textwidth}{!}{
        \begin{tabular}{c|c|c|c|c|c|c|c|c}
            \toprule
            \makecell{\textbf{Incremental}\\\textbf{Dataset}} & \makecell{\textbf{Data}\\\textbf{Amount}} &
            \textbf{SEED\textsuperscript{I}} &\textbf{MMB\textsuperscript{Dev}} & \textbf{MME}
            & \textbf{GQA} & \textbf{VizWiz} & \textbf{VQAv2} & \textbf{ScienceQA} \\
            \midrule
            None (baseline) & - & 66.04 & 66.66 & 1475/290(1765) & 57.27 & 49.07 & 77.56 & 70.67/68.27 \\
            \midrule
             LRV~\cite{liu2023mitigating} & 320k & 67.71 & 67.61 & 1462/304(1766) & 58.23 & 50.41 & 78.22 & 71.80/69.46 \\
             SVIT~\cite{DBLP:journals/corr/abs-2307-04087} & 1.5M & 68.22 &66.66 & 1415/282(1697) & \textbf{58.55} & 46.96 & \textbf{78.91} & 70.34/68.72 \\
             \midrule
             DataEngine, Round1 & 80k & 67.72 & 67.47 & 1484/270(1754) & 57.33 & 51.26 & 77.72 & 72.41/71.29 \\
            DataEngine, Round2 & 170k & 68.30 & \textbf{67.61} & 1486/286(1772) & 57.74 & 48.95 & 78.18 & 72.37/71.19 \\
            \midrule
            \multirow{2}{*}{DataEngine, Round3}
             & \multirow{2}{*}{220k} & \textbf{68.57} & 67.18 & \textbf{1511/303(1814)} & 58.02 & \textbf{52.90} & \Gape{\makecell{78.18\\ }} & \textbf{73.17/71.15} \\
             & & \textcolor{red}{$\uparrow$\textit{2.53}} & \textcolor{red}{$\uparrow${\textit{0.52}}} & \textcolor{red}{$\uparrow${\textit{36/13(49)}}} & \textcolor{red}{$\uparrow${\textit{0.75}}} & \textcolor{red}{$\uparrow${\textit{3.83}}} & \textcolor{red}{$\uparrow${\textit{0.62}}} & \textcolor{red}{$\uparrow${\textit{2.50/2.88}}} \\
            \bottomrule
        \end{tabular}
    }
    \label{table_res1}
\end{table*}

\subsection{Prompt Construction} \label{sec: Prompt Construction}
After the bad case pool is established, we construct delicate prompts for further data generation, which is composed of two key components: (1) Detailed image information described in language and (2) Well-demonstrated question type. 
Each component plays a crucial role in the generation of high-quality data.

Firstly, for detailed image information, we randomly choose images from Visual Genome (VG) Dataset~\cite{DBLP:journals/ijcv/KrishnaZGJHKCKL17}, which provides a range of annotations for each image, such as Region Descriptions, Object Instances, and Object Attributes, \textit{et al.}. 
Among those, Region Descriptions, which contain multiple instances along with detailed descriptions of objects as well as their position coordinates, are extracted as Image Information. 
Moreover, for the Text Understanding task, we employ PaddleOCR~\cite{DBLP:journals/corr/abs-2009-09941} to extract texts that appear in images, which serves as an auxiliary supplement to the image information.

Secondly, to clearly demonstrate the specific question type expected to be generated from GPT-4, a clear definition of this question type and in-context learning examples are needed. 
To ensure the effective selection of appropriate in-context examples and to align the generated data with the model's shortcomings, we introduce the Adaptive Bad-case Sampling (ABS) strategy. 
It comprises two steps: (1) A question type is chosen randomly based on an adaptive sampling ratio. The sampling ratio for each question type is based on the corresponding evaluation dimensions scores, which is represented by $r_{i} = \sqrt{1 - a_{i}}$, where $r_{i}$ is the sampling ratio of $i^{th}$ question type and $a_{i}$ denotes the accuracy attained by MLLM in $i^{th}$ question type.
(2) Ten instances of this question type are then randomly selected from the bad case pool to serve as in-context examples, assisting GPT-4 in understanding the required question type effectively and guiding GPT-4 to generate valuable data.
ABS ensures the flexibility and adaptability of data generation and more data will be generated for dimensions where the model's performance is poorer.
Consequently, ABS effectively guides the generation of data to address the weaknesses of the model.

\subsection{Data Generation}

With the prompt constructed, we utilize \emph{gpt-4-1106-preview} version of GPT-4 to generate instruct fine-tuning data.
During data generation, GPT-4 is prompted to construct diverse and complex questions, along with corresponding accurate Direct Answers~(DA) based on the provided information. 
In addition to direct answers, GPT-4 is also instructed to reformat questions into a Multiple-Choice format~(MC), comprising four options (A, B, C, and D) and a correct answer. 
Meanwhile, to further enhance the data diversity, we randomly shuffle options in each multi-choice question. Example generated data are shown in Fig.~\ref{fig:ability}.

\begin{table*}[t]
    \caption{Experiments on MiniGPT4-v2. The None~(baseline) row indicates the baseline model results.}
    \centering
    \resizebox{0.89\textwidth}{!}{
        \begin{tabular}{c|c|c|c|c|c|c}
            \toprule
            \textbf{Incremental Dataset}
            & 
            \textbf{Data Amount}
            &
            \textbf{SEED\textsuperscript{I}} &\textbf{MMB\textsuperscript{Dev}} & \textbf{OK-VQA}
            & \textbf{VizWiz} & \textbf{VSR} \\
            \midrule
            None (baseline) & - & 49.21 & 38.83 & 56.03 & 53.08 & 61.37 \\
            \midrule
             LRV~\cite{liu2023mitigating} & 320k & 49.24 & 40.72 & 56.66 & 54.30 & 61.78 \\
             SVIT~\cite{DBLP:journals/corr/abs-2307-04087} & 1.5M & 49.75 & 43.64 & 56.95 & 54.31 & 58.67 \\
             \midrule
             DataEngine, Round1 & 100k & 61.84 & 51.80 & 56.76 & 53.50 & 62.09 \\
            DataEngine, Round2 & 180k & 63.80 & \textbf{53.43} & \textbf{57.05} & 53.62 & \textbf{62.52} \\
            \midrule
            \multirow{2}{*}{DataEngine, Round3} & \multirow{2}{*}{270k} & \textbf{63.83} & 52.92 & 56.87 & \textbf{54.39} & 62.43 \\
            & & \textcolor{red}{$\uparrow${\textit{14.62}}} & \textcolor{red}{$\uparrow${\textit{14.09}}} & \textcolor{red}{$\uparrow${\textit{0.74}}} & \textcolor{red}{$\uparrow${\textit{1.31}}} & \textcolor{red}{$\uparrow${\textit{1.06}}} \\
            \bottomrule
        \end{tabular}
    }
    
    \label{table_res2}
\end{table*}

\begin{figure}[t]
    \centering
    \includegraphics[width=0.90\linewidth]{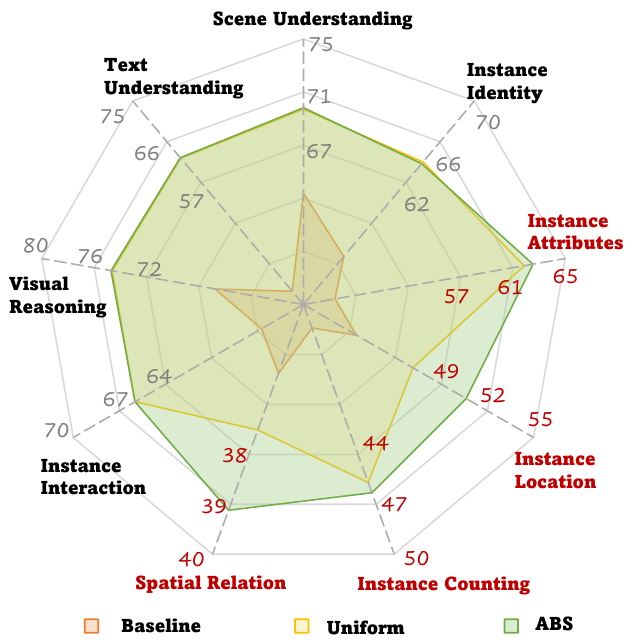}
    \caption{Comparison between uniform sampling and Adaptive Bad-case Sampling~(ABS). Weak capabilities of the baseline model are highlighted.}
    \label{fig:ablation_abs}
\end{figure}

\section{Experiments}


\begin{table}[t]
    \caption{Experiments of different synthetic data formats on LLaVA-1.5. \emph{DA} refers to direct answer and \emph{MC} refers to multiple choice answer.}
   \centering
   \resizebox{0.47\textwidth}{!}{
       \begin{tabular}{c|c|c|c|c}
           \toprule
           \makecell{\textbf{Incremental}\\\textbf{Dataset}} & \textbf{Data Format} &
           \textbf{SEED\textsuperscript{I}} &\textbf{MMB\textsuperscript{Dev}}
           & \textbf{GQA}\\
           \midrule
           None (baseline) & - & 66.04 & 66.66 & 57.27 \\
           \midrule
           \multirow{2}{*}{DataEngine, Round1} &
            {DA} & 66.21 & 67.26 & 58.21 \\
            & DA + MC & 67.72 & 67.47 & 57.33\\
            \midrule
            \multirow{2}{*}{DataEngine, Round2} &
            {DA} & 67.25 & 67.18 & 58.19 \\
            & DA + MC & 68.30 & \textbf{67.61} & 57.74\\
           \midrule
           \multirow{2}{*}{DataEngine, Round3} &
            {DA} & 67.33 & 66.15 & \textbf{58.60} \\
            & DA + MC & \textbf{68.57} & 67.18 & 58.02\\
           \bottomrule
       \end{tabular}
   }
   
   \label{tab:abl3}
\end{table}

\subsection{Implementation Details}

We adopt three mainstream, open-source MLLMs in experiments: LLaVA-1.5~\cite{DBLP:journals/corr/abs-2310-03744}, MiniGPT4-v2~\cite{DBLP:journals/corr/abs-2310-09478}, and MiniGPT-4~\cite{DBLP:journals/corr/abs-2304-10592}. We incorporate these MLLMs into MLLM-DataEngine for iterative refinement. Specifically, we use both the original instruct-tuning data and MLLM-DataEngine generated incremental data in the instruct tuning of each iteration. The refined model is evaluated using various downstream benchmarks for a comprehensive assessment. Details can be found in supplementary materials.

\subsection{Main Results}
\label{sec:main_results}


In this section, we conduct multiple rounds of model refinement using our proposed MLLM-DataEngine to validate its effectiveness. In each round of refinement in MLLM-DataEngine, we combine the incremental data generated by MLLM-DataEngine~(data generated in the current and the previous rounds) with the original instruction fine-tuning data for fine-tuning.

Experiments are conducted on LLaVA-1.5 and MiniGPT4-v2, whose results are demonstrated in Table~\ref{table_res1} and Table~\ref{table_res2}, respectively. The \emph{None~(baseline)} row indicates baseline model results without incremental dataset. SEED\textsuperscript{I} refers to SEED-Bench~(Image). MMB\textsuperscript{Dev} refers to MMBench Dev. MME score stands for Perception/Cognition~(total score). ScienceQA score stands for Overall score/Image score. Besides, we compare our proposed MLLM-DataEngine with two comparison methods: LRV~\cite{liu2023mitigating} and SVIT~\cite{DBLP:journals/corr/abs-2307-04087}, which also seek to enhance MLLMs by scaling up instruction fine-tuning data. From the results, we can see that: \textbf{(1) MLLM-DataEngine can effectively and iteratively enhance model performance.} Results clearly show that the model performance increases in each round of MLLM-DataEngine across various benchmarks as the generated data amount grows. For instance, three rounds of refinement on LLaVA-1.5 result in improvements of 2.53\% for SEED-Bench, 0.52\% for MMBench Dev, and a total score increase of 49 for MME when compared to baseline. Additionally, superior results are also achieved on traditional VQA benchmarks under MLLM-DataEngine refinement. \textbf{(2) MLLM-DataEngine presents a more effective data generation strategy.} MLLM-DataEngine achieves significantly better improvements by using a more targeted but lesser volume of data compared with LRV and SVIT, which produce 320k and 1.5M instruction fine-tuning data for model enhancement respectively. This indicates the superiority of MLLM-DataEngine in terms of data generation strategy, as well as the effectiveness of the underlying iterative refinement philosophy. \textbf{(3) The improvement brought by data scaling is not unlimited.} Comparing results between 3 rounds of refinement in MLLM-DataEngine, it can be seen that: In the beginning, there was a significant improvement through data scaling. However, as the volume of data grows to a certain extent, The trade-off between data scaling and model improvements is becoming increasingly smaller. For example, MiniGPT4-v2's third round of refinement demonstrated significantly less improvement compared to the first and second rounds. We speculate that further improvement is constrained due to potential model limitations such as input resolution, visual feature granularity, and so on.

\subsection{Ablation Studies}\label{sec:abl}

\subsubsection{Comparison between ABS and Uniform Sampling}

To further validate the ability of Adaptive Bad-case Sampling (ABS) to enhance model weakness, we compared its effects with uniform sampling. For comparison, we directly sample from all generated data, using both ABS and uniform sampling, with each method fixedly sampling 90K data. We fine-tune the baseline MiniGPT4-v2 model using data obtained from those two sampling strategies separately. Results on SEED-Bench are shown in Fig.~\ref{fig:ablation_abs}~(detailed result in supplementary). Results demonstrate that for the weaknesses of the baseline model, our proposed Adaptive Bad-case Sampling (ABS) can provide a more targeted enhancement.



\subsubsection{Effects of Different Instruction Tuning Data Formats}\label{sec:abl3}

To explore the effects of different data formats, we experiment with different formats on LLaVA-1.5, and results are shown in Table~\ref{tab:abl3}. \emph{DA} refers to direct answer and \emph{MC} refers to multiple choices question. Firstly, regardless of the format, significant performance improvements can be achieved by generated data. For example, the three-round DA also has a significant improvement compared to the baseline model. Secondly, the choice of data format influences performance across different benchmarks. For example, the multiple choices~(MC) leads to substantial improvements on challenging multiple-choice benchmarks such as SEED and MMBench when compared to Direct Answer~(DA); however, on reasoning VQA benchmarks, using only DA yields the best results~(58.60\% accuracy on GQA).

\begin{table}[t]
    \caption{MiniGPT4-v2~(A) $\to$ MiniGPT4~(B).}
    \vspace{-5pt}
    \centering
    \resizebox{0.40\textwidth}{!}{
        \begin{tabular}{c|c|c}
            \toprule
             \textbf{Incremental Dataset} & \textbf{SEED\textsuperscript{I}} & \textbf{MMB\textsuperscript{Dev}} \\
             \midrule
             None, baseline & 21.30 & 23.00 \\
             \midrule
             MiniGPT4-v2, Round1 & 54.31 & 38.91 \\
             MiniGPT4-v2, Round2 & 56.80 & 46.30 \\
             \midrule
             MiniGPT4-v2, Round3 & \textbf{58.12} & \textbf{49.31} \\
             \bottomrule
        \end{tabular}   
    }
    
    \label{tab:abl1_1}
\end{table}

\begin{table}[t]
    \caption{LLaVA-1.5~(A) $\to$ MiniGPT4~(B).}
    \vspace{-5pt}
    \centering
    \resizebox{0.40\textwidth}{!}{
        \begin{tabular}{c|c|c}
            \toprule
             \textbf{Incremental Dataset} & \textbf{SEED\textsuperscript{I}} & \textbf{MMB\textsuperscript{Dev}} \\
             \midrule
             None, baseline & 21.30 & 23.00 \\
             \midrule
             LLaVA-1.5, Round1 & 51.96 & 39.00 \\
             LLaVA-1.5, Round2 & 57.11 & \textbf{48.45} \\
             \midrule
             LLaVA-1.5, Round3 & \textbf{58.34} & 45.79 \\
             \bottomrule
        \end{tabular}
    }
    
    \label{tab:abl1_2}
\end{table}

\begin{table}[t]
    \caption{LLaVA-1.5~(A) $\to$ MiniGPT4-v2~(B).}
    \vspace{-5pt}
    \centering
    \resizebox{0.40\textwidth}{!}{
        \begin{tabular}{c|c|c}
            \toprule
             \textbf{Incremental Dataset} & \textbf{SEED\textsuperscript{I}} & \textbf{MMB\textsuperscript{Dev}} \\
             \midrule
             None, baseline & 49.21 & 38.83 \\
             \midrule
             LLaVA-1.5, Round1 & 59.81 & 48.71 \\
             LLaVA-1.5, Round2 & 62.72 & {51.67} \\
             \midrule
             LLaVA-1.5, Round3 & \textbf{63.09} & \textbf{52.14} \\
             \bottomrule
        \end{tabular}
    }
    
    \label{tab:abl1_3}
\end{table}

\begin{table}[t]
    \caption{MiniGPT4-v2~(A) $\to$ LLaVA-1.5~(B).}
    \vspace{-5pt}
    \centering
    \resizebox{0.40\textwidth}{!}{
        \begin{tabular}{c|c|c}
            \toprule
             \textbf{Incremental Dataset} & \textbf{SEED\textsuperscript{I}} & \textbf{MMB\textsuperscript{Dev}} \\
             \midrule
             None, baseline & 66.04 & 66.66 \\
             \midrule
             MiniGPT4-v2, Round1 & 67.40 & 66.23 \\
             MiniGPT4-v2, Round2 & 67.95 & {66.87} \\
             \midrule
             MiniGPT4-v2, Round3 & \textbf{68.46} & \textbf{67.04} \\
             \bottomrule
        \end{tabular}
    }
    
    \label{tab:abl1_4}
\end{table}

\subsubsection{Generalization Ability of Generated Data}\label{sec:abl1}

Lastly, we explore the generalization capability of the data generated by MLLM-DataEngine. More specifically, we address the question, ``Is the data generated for Model A also beneficial for Model B?''. 
We investigate this by using model A's MLLM-DataEngine generated data to instruction fine-tune model B and examining subsequent performance improvements~(represented as A$\to$B). Table~\ref{tab:abl1_1} and Table~\ref{tab:abl1_2} demonstrate the results of fine-tuning MiniGPT-4~(model B) using the data generated with MiniGPT4-v2 and LLaVA-1.5 in the MLLM-DataEngine loop, respectively~(model A). 
MLLM-DataEngine generated data for MiniGPT4-v2 and LLaVA-1.5 achieves $36.82\%$/$26.31\%$ and $37.04\%$/$22.79\%$ improvement in MiniGPT4, respectively. We further extend this investigation with cross-validation on two strong MLLMs, MiniGPT4-v2 and LLaVA-1.5. Results are shown in Table~\ref{tab:abl1_3} and Table~\ref{tab:abl1_4}, which also exhibit performance improvements, confirming that the data generated by MLLM-DataEngine is not only model-specific but also broadly applicable across different MLLMs.

\section{Conclusion}

This paper presents MLLM-DataEngine, a framework for generating high-quality, targeted instruction fine-tuning data, addressing model weaknesses, and forming a closed training loop for large multi-modal models. We hope this approach will advance data construction on multi-modal research.

\bibliographystyle{IEEEbib}
\bibliography{icme2025references}

\end{document}


\title{Supplementary Materials for ``MLLM-DataEngine: Closing the Loop of Multimodal Instruction Tuning Data Generation"}

\maketitle

\begin{figure}[t]
    \centering
    \includegraphics[width=0.95\linewidth]{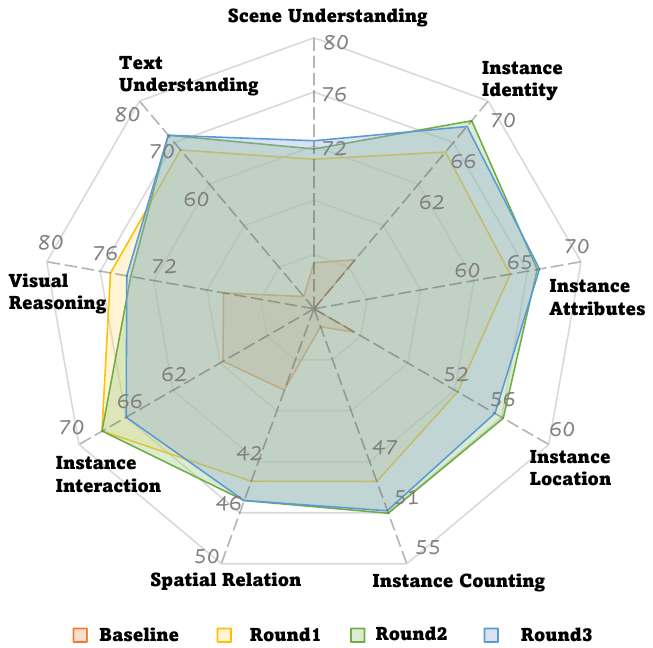}
    \caption{Results of each capability of MiniGPT4-v2 during the refinement process of MLLM-DataEngine.}
    \label{fig:seed_detail}
\end{figure}

\begin{table*}[t]
    \centering
    \caption{SEED-Bench results of LLaVA-1.5 during MLLM-DataEngine refinement.}
    \label{supp:table1}
    \resizebox{1.0\textwidth}{!}{
        \begin{tabular}{c|c|c|c|c|c|c|c|c|c|c}
            \toprule
            \makecell{\textbf{Incremental}\\\textbf{Dataset}} & \makecell{\textbf{Scene}\\\textbf{Understanding}} &
            \makecell{\textbf{Instance}\\\textbf{Identity}} & \makecell{\textbf{Instance}\\\textbf{Attributes}} & \makecell{\textbf{Instance}\\\textbf{Location}}
            & \makecell{\textbf{Instance}\\\textbf{Counting}} & \makecell{\textbf{Spatial}\\\textbf{Relation}} & \makecell{\textbf{Instance}\\\textbf{Interaction}} & \makecell{\textbf{Visual}\\\textbf{Reasoning}} & \makecell{\textbf{Text}\\\textbf{Understading}} &
            \textbf{Overall} \\
            \midrule
            None (baseline) & 74.89 & 68.32 & 66.87 & 60.94 & 57.42 & 49.92 & 73.20 & 76.13 & 27.06 & 66.04 \\
            \midrule
             Data-Engine, round1 & 73.75 & 69.52 & 70.02 & 59.30 & 59.95 & 54.34 & 71.13 & 76.13 & 47.06 & 67.22 \\
            Data-Engine, round2 & 74.57 & 71.00 & 71.67 & 60.63 & 58.52 & 53.42 & 69.07 & 75.83 & 67.06 & 68.30 \\
            \midrule
            Data-Engine, round3 & 74.32 & 70.51 & 72.10 & 62.17 & 58.19 & 55.10 & 72.16 & 77.04 & 60.00 & 68.57 \\
            \bottomrule
        \end{tabular}
    }
\end{table*}

\begin{table*}[t]
    \centering
    \caption{SEED-Bench results of MiniGPT4-v2 during MLLM-DataEngine refinement.}
    \label{supp:table2}
    \resizebox{1.0\textwidth}{!}{
        \begin{tabular}{c|c|c|c|c|c|c|c|c|c|c}
            \toprule
            \makecell{\textbf{Incremental}\\\textbf{Dataset}} & \makecell{\textbf{Scene}\\\textbf{Understanding}} &
            \makecell{\textbf{Instance}\\\textbf{Identity}} & \makecell{\textbf{Instance}\\\textbf{Attributes}} & \makecell{\textbf{Instance}\\\textbf{Location}}
            & \makecell{\textbf{Instance}\\\textbf{Counting}} & \makecell{\textbf{Spatial}\\\textbf{Relation}} & \makecell{\textbf{Instance}\\\textbf{Interaction}} & \makecell{\textbf{Visual}\\\textbf{Reasoning}} & \makecell{\textbf{Text}\\\textbf{Understading}} &
            \textbf{Overall} \\
            \midrule
            None (baseline) & 63.39 & 54.72 & 47.41 & 43.46 & 36.41 & 36.38 & 57.73 & 66.77 & 32.94 & 49.66 \\
            \midrule
             Data-Engine, round1 & 71.03 & 65.10 & 64.70 & 52.25 & 48.55 & 43.53 & 68.04 & 75.23 & 68.24 & 61.84 \\
            Data-Engine, round2 & 71.79 & 68.10 & 66.72 & 56.13 & 51.04 & 45.05 & 68.04 & 73.72 & 71.76 & 63.80 \\
            \midrule
            Data-Engine, round3 & 72.39 & 67.56 & 66.90 & 55.42 & 50.84 & 45.05 & 65.98 & 74.02 & 71.76 & 63.83 \\
            \bottomrule
        \end{tabular}
    }
\end{table*}


In Section~\ref{sec:1}, we provide prompts used in MLLM-DataEngine data generation. In Section~\ref{sec:2}, we provide implementation details. In Section~\ref{sec:3}, we provide anylysis on MLLM-DataEngine generated data. In Section~\ref{sec:4}, we provide detailed results during each optimization ieration of MLLM-DataEngine. In Section~\ref{sec:5}, we provide eaxmple MLLM-DataEngine generated data.

\section{Prompts Used in MLLM-DataEngine}
\label{sec:1}

\begin{figure*}[ht]
    \centering
    \includegraphics[width=1.0\textwidth]{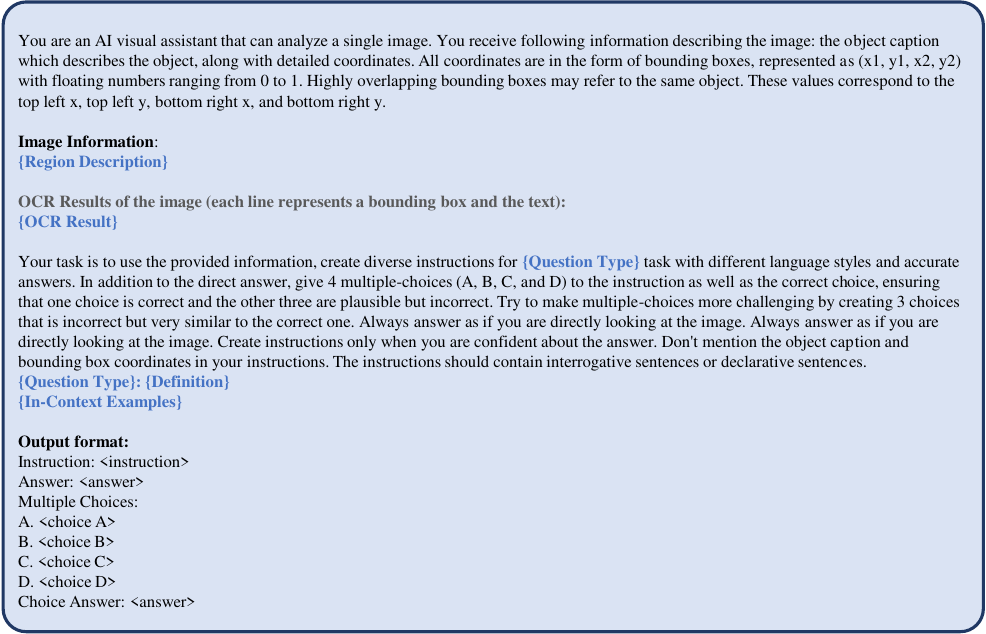}
    \caption{Prompt Template for data generation in MLLM-DataEngine.}
    \label{fig:prompt}
\end{figure*}

Fig.~\ref{fig:prompt} is the prompt template used in MLLM-DataEngine when generating data with GPT-4. Among it, Region Description is extracted from Visual Genome (VG) Dataset. OCR Result is the output of PaddleOCR when generating Text Understanding questions. Question Types and its Definition are acquire from SEED Benchmark. In-Context Examples are randomly selected from Bad Case Pool of MLLMs in that question type. 

\begin{figure*}[t]
    \centering
    \subfloat[Visual Reasoning]{\includegraphics[scale=0.2]{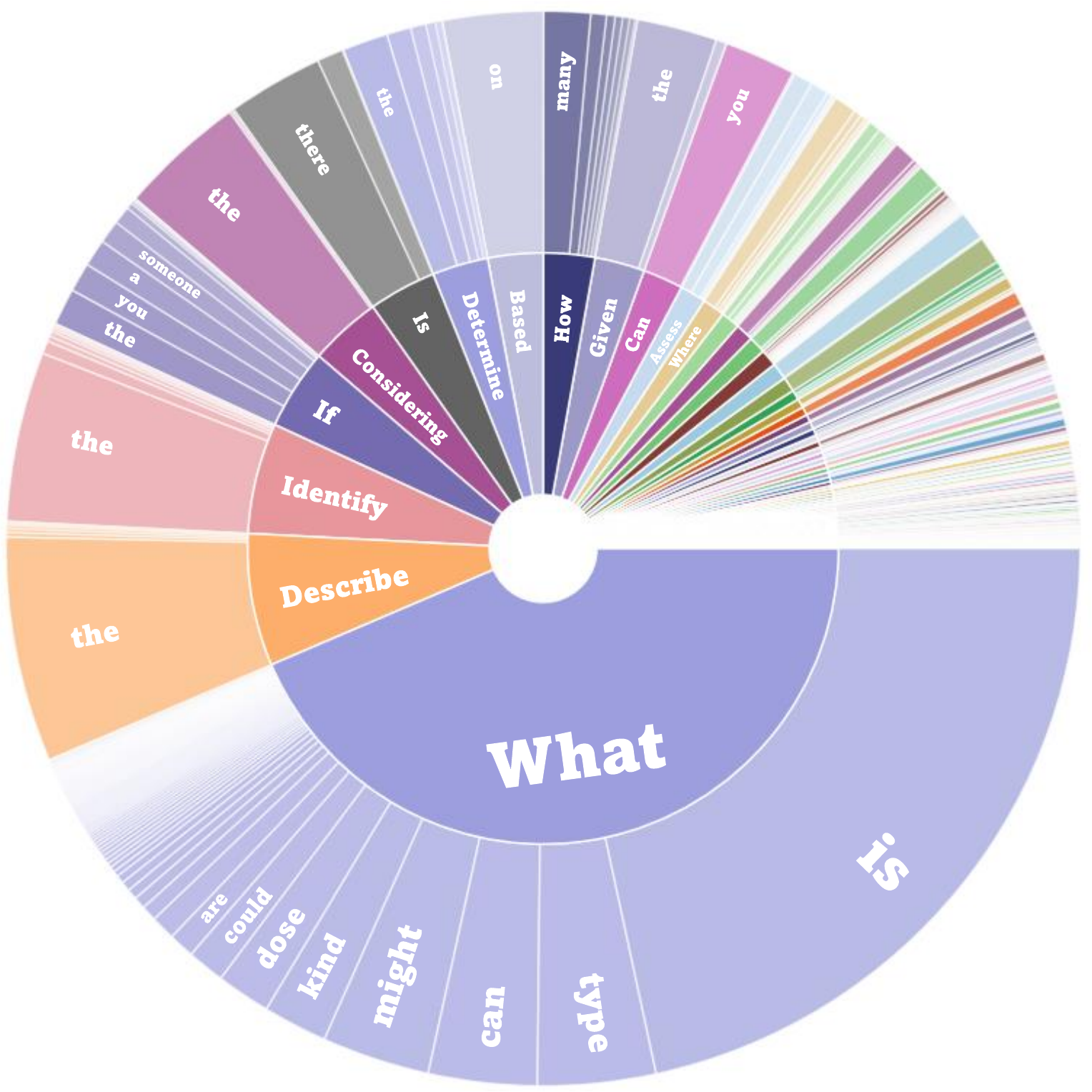}}
    \label{fig:word1}
    \hfill
    \subfloat[Spatial Relation]{\includegraphics[scale=0.2]{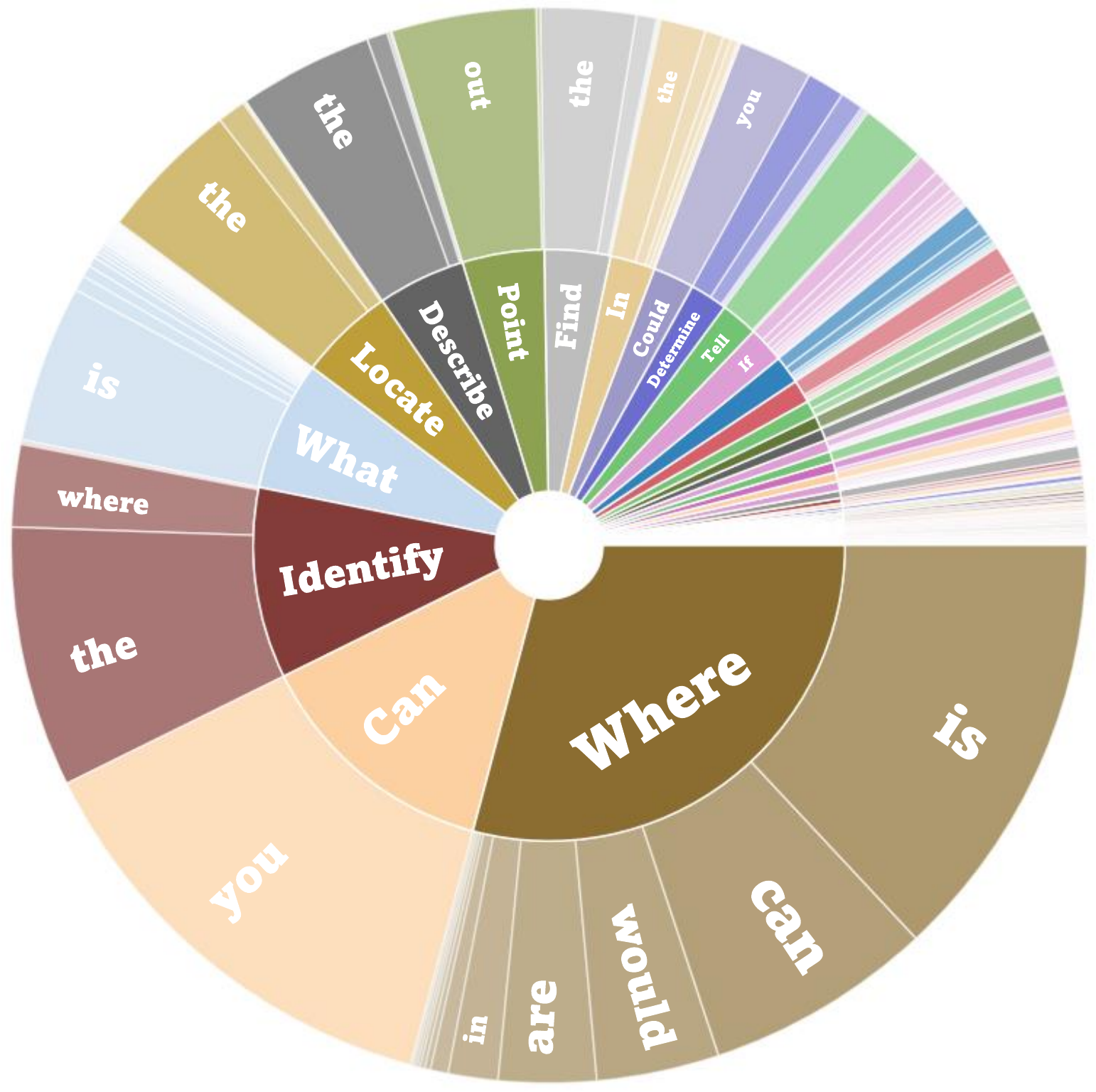}}
    \label{fig:word2}
    \hfill
    \subfloat[Object Counting]{\includegraphics[scale=0.2]{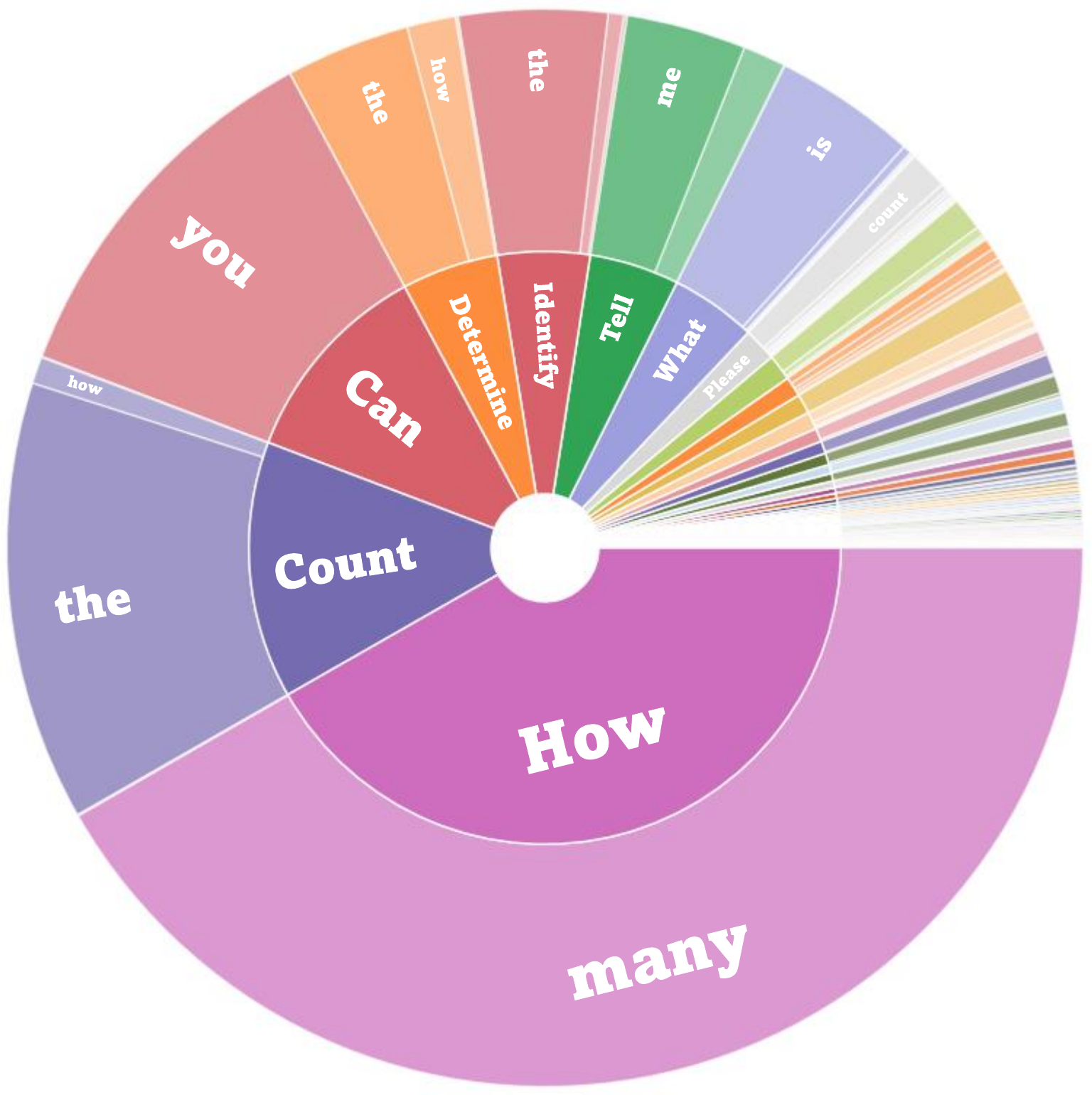}}
    \label{fig:word3}
    \caption{Words distributions analysis. Generated instruct data aligns closely to the given question type.}
    \label{fig:word_distribution}
\end{figure*}

\section{Implementation Details}
\label{sec:2}

\subsection{LLaVA-1.5} is the SOTA~(state-of-the-art) open-source MLLM with strong capabilities in image content recognition and understanding. The instruction fine-tuning data covering open-knowledge VQA~\cite{DBLP:conf/cvpr/GoyalKSBP17, DBLP:conf/cvpr/MarinoRFM19, DBLP:conf/eccv/SchwenkKCMM22, DBLP:conf/cvpr/HudsonM19}, OCR~\cite{DBLP:conf/icdar/0001SSC19, DBLP:conf/eccv/SidorovHRS20}, conversation~\cite{DBLP:journals/corr/abs-2304-08485, sharegpt}, and region-level VQA~\cite{DBLP:journals/ijcv/KrishnaZGJHKCKL17, DBLP:conf/cvpr/MaoHTCY016, DBLP:conf/emnlp/KazemzadehOMB14}, leading to a total of 665k instruction fine-tuning data. Evaluations are conducted on comprehensive multi-dimensional benchmarks~(SEED-Bench~\cite{DBLP:journals/corr/abs-2307-16125}, MMBench~\cite{DBLP:journals/corr/abs-2307-06281}, MME~\cite{DBLP:journals/corr/abs-2306-13394}) and conventional VQA datasets~(VQAv2, GQA, VizWiz~\cite{DBLP:conf/cvpr/Gurari0SGLGLB18}, ScienceQA~\cite{DBLP:conf/nips/LuMX0CZTCK22}). As for the experimental setting, we adopt \emph{LLaVA-1.5-7b-lora} version of LLaVA-1.5, where the model is fine-tuned for 1 epoch using LoRA~\cite{DBLP:conf/iclr/HuSWALWWC22} with \emph{rank} set to 128 and \emph{alpha} set to 256, following its original setting. The cosine learning rate scheduler is adopted with an initial learning rate set to $1e^{-4}$ and a warm-up ratio set to $0.03$. The model is fine-tuned using 8$\times$A100 GPUs for about 20 hours.

\subsection{MiniGPT4-v2}
MiniGPT4-v2 is another strong-performed MLLM that is fine-tuned using abundant multi-task instruction data. The multimodel instruction fine-tuning data of MiniGPT4-v2 consists of multiple tasks: image captioning~\cite{DBLP:conf/eccv/LinMBHPRDZ14, DBLP:conf/eccv/SidorovHRS20}, referring expression comprehension/generation~\cite{DBLP:conf/eccv/YuPYBB16, DBLP:conf/cvpr/MaoHTCY016, DBLP:conf/emnlp/KazemzadehOMB14}, VQA~\cite{DBLP:conf/cvpr/GoyalKSBP17, DBLP:conf/cvpr/MarinoRFM19, DBLP:conf/eccv/SchwenkKCMM22, DBLP:conf/cvpr/HudsonM19, DBLP:conf/icdar/0001SSC19}, and multimodel conversation~\cite{DBLP:conf/iccv/PlummerWCCHL15, DBLP:journals/corr/abs-2304-08485}. During evaluation, except from three VQA datasets used in the original setting~(OKVQA~\cite{DBLP:conf/cvpr/MarinoRFM19}, VizWiz~\cite{DBLP:conf/cvpr/Gurari0SGLGLB18}, VSR~\cite{liu2023visual}), we carry out evaluations on SEED-Bench and MMBench. As for the experimental setting, we follow the original setting and fine-tune the model for 10/20/30 epochs with 1k iterations per epoch as the amount of data grows in each round of MLLM-DataEngine refinement. LoRA with \emph{rank} set to 16 and \emph{alpha} set to 64 is used. The cosine learning rate scheduler is adopted with an initial learning rate set to $1e^{-5}$ and a warm-up step set to 1k. The model is fine-tuned using 8$\times$A100 GPUs for about 10 hours.

\subsection{MiniGPT4} MiniGPT4 the preliminary version of MiniGPT4-v2, has basic image comprehension and instruction-following abilities but is inferior across various capability dimensions. The model is fine-tuned using 3.5k filtered image captioning data~\cite{DBLP:journals/corr/abs-2304-10592}. SEED-Bench and MMBench are adopted as evaluation benchmarks. In terms of the experimental setup, we employ LoRA for one epoch with \emph{rank} set to 16 and \emph{alpha} to 64. The model uses a cosine learning rate scheduler with an initial rate of $1e^{-5}$ and 1k warm-up step, fine-tuned on 8$\times$A100 GPUs over approximately 2-3 hours.


\section{Data Statistics and Analysis}
\label{sec:3}

 The average lengths of the generated questions and answers are 67.60 and 76.81 characters, respectively and questions contain both interrogative and declarative sentences. To validate whether the generated instructions align with the given question type, we further analyze the distribution of the first and second words in the generated questions. As shown in Fig.~\ref{fig:word_distribution}, the inner circle of the diagram represents the first word of the question, while the outer circle signifies the second word. It can be concluded that the generated data accurately reflects the given capability dimensions. For instance, Visual Reasoning requires an accurate comprehension of image content and context, hence questions like ``What can'', ``What might'', and ``What could'' are prevalent.

\section{Detailed MLLM-DataEngine Results}
\label{sec:4}

Detailed SEED-Bench results in each round of LLaVA-1.5 and MiniGPT4-v2 are shown in Table~\ref{supp:table1} and Table~\ref{supp:table2}, respectively. Results of each capability of MiniGPT4-v2 during
the refinement process of MLLM-DataEngine is illustrated in Fig.~\ref{fig:seed_detail}.

\section{Quality Examples in MLLM-DataEngine Generated Data}
\label{sec:5}

\begin{figure*}[t]
    \centering
    \includegraphics[width=0.85\textwidth]{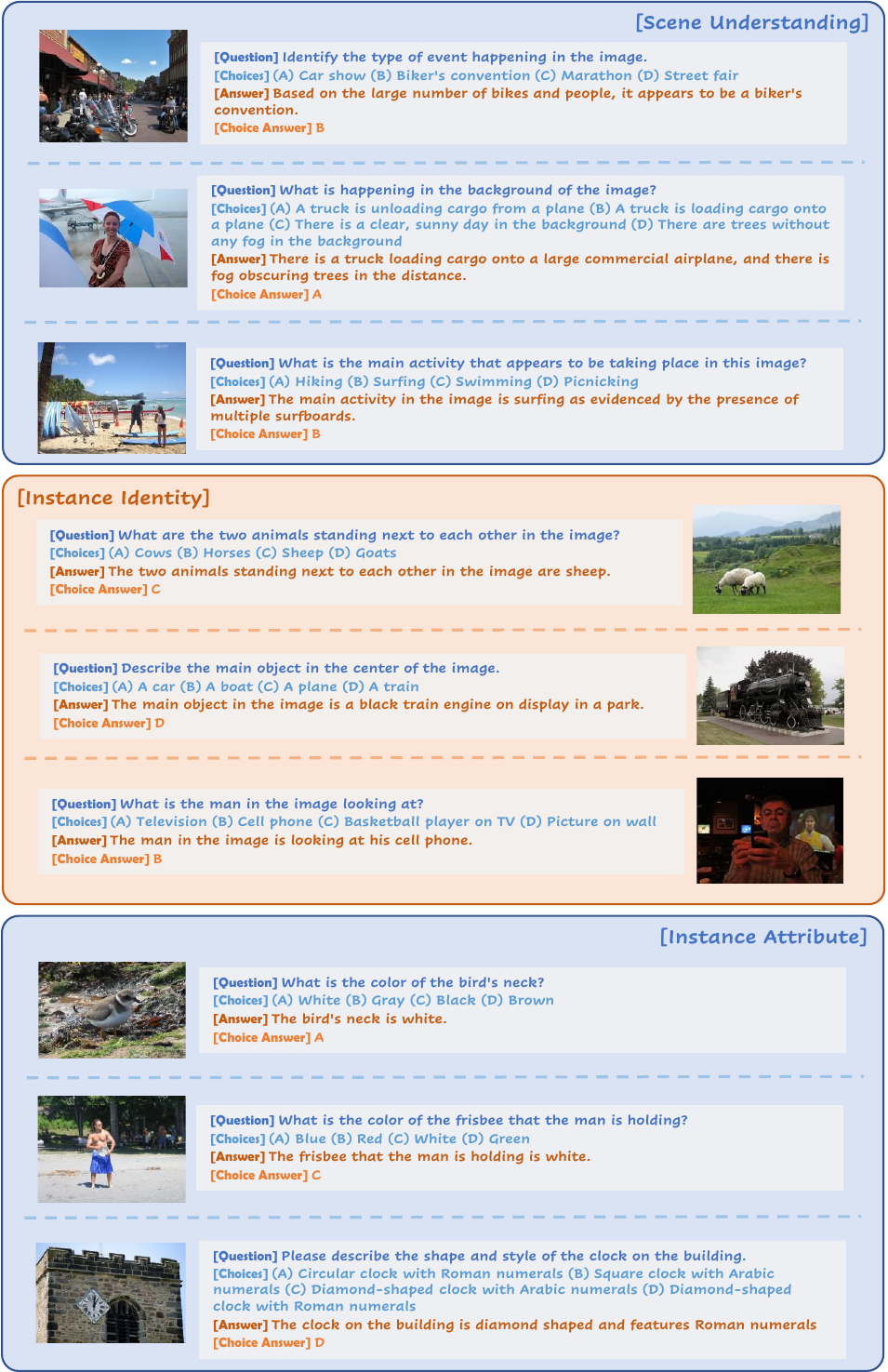}
    \caption{Examples of MLLM-DataEngine generated data in Scene Understanding, Instance Identity, and Instance Attribute.}
    \label{fig:Showcase1}
\end{figure*}

\begin{figure*}[t]
    \centering
    \includegraphics[width=0.85\textwidth]{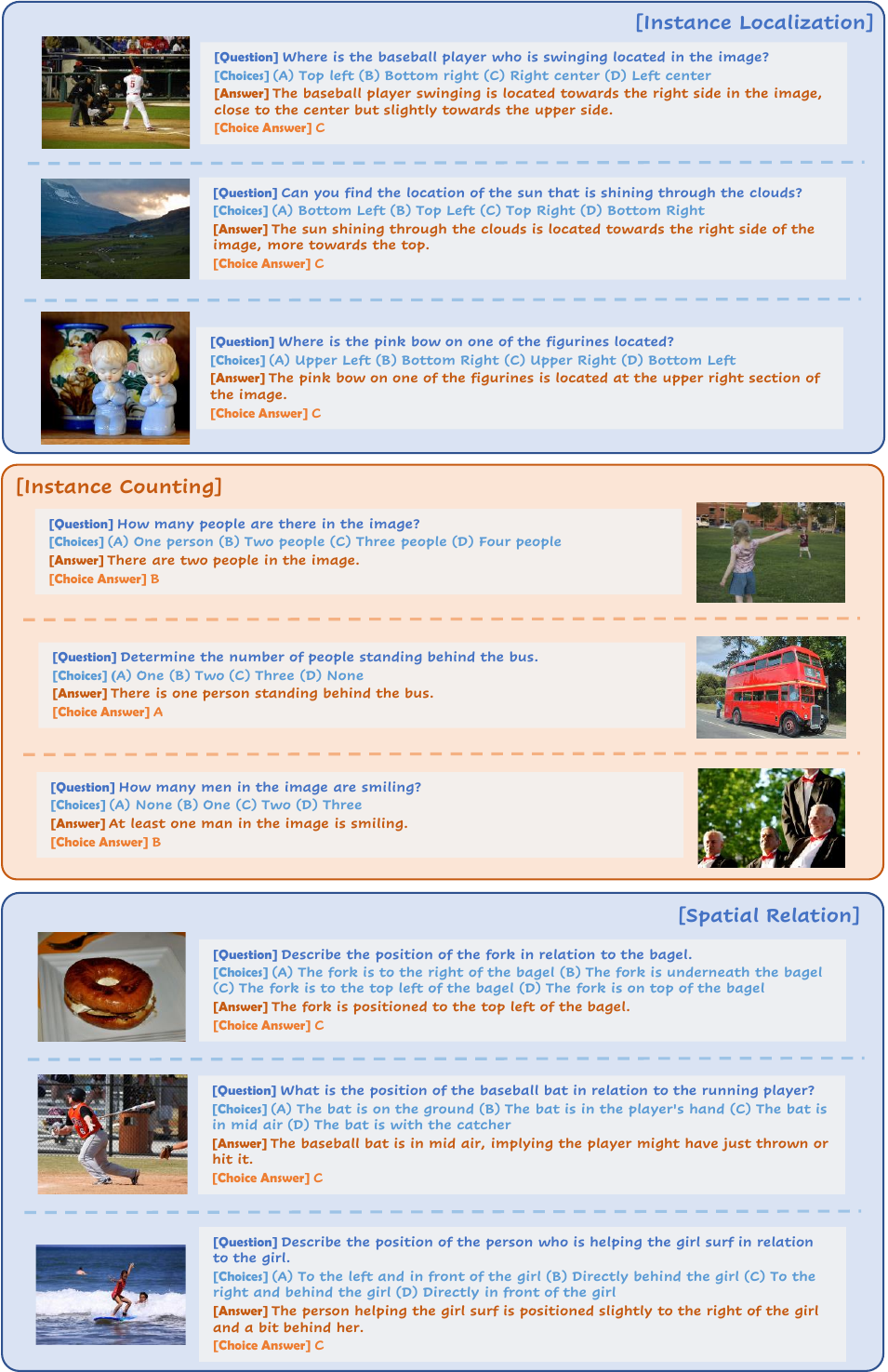}
    \caption{Examples of MLLM-DataEngine generated data in Instance Localization, Instance Counting, and Spatial Relation.}
    \label{fig:Showcase2}
\end{figure*}

\begin{figure*}[t]
    \centering
    \includegraphics[width=0.85\textwidth]{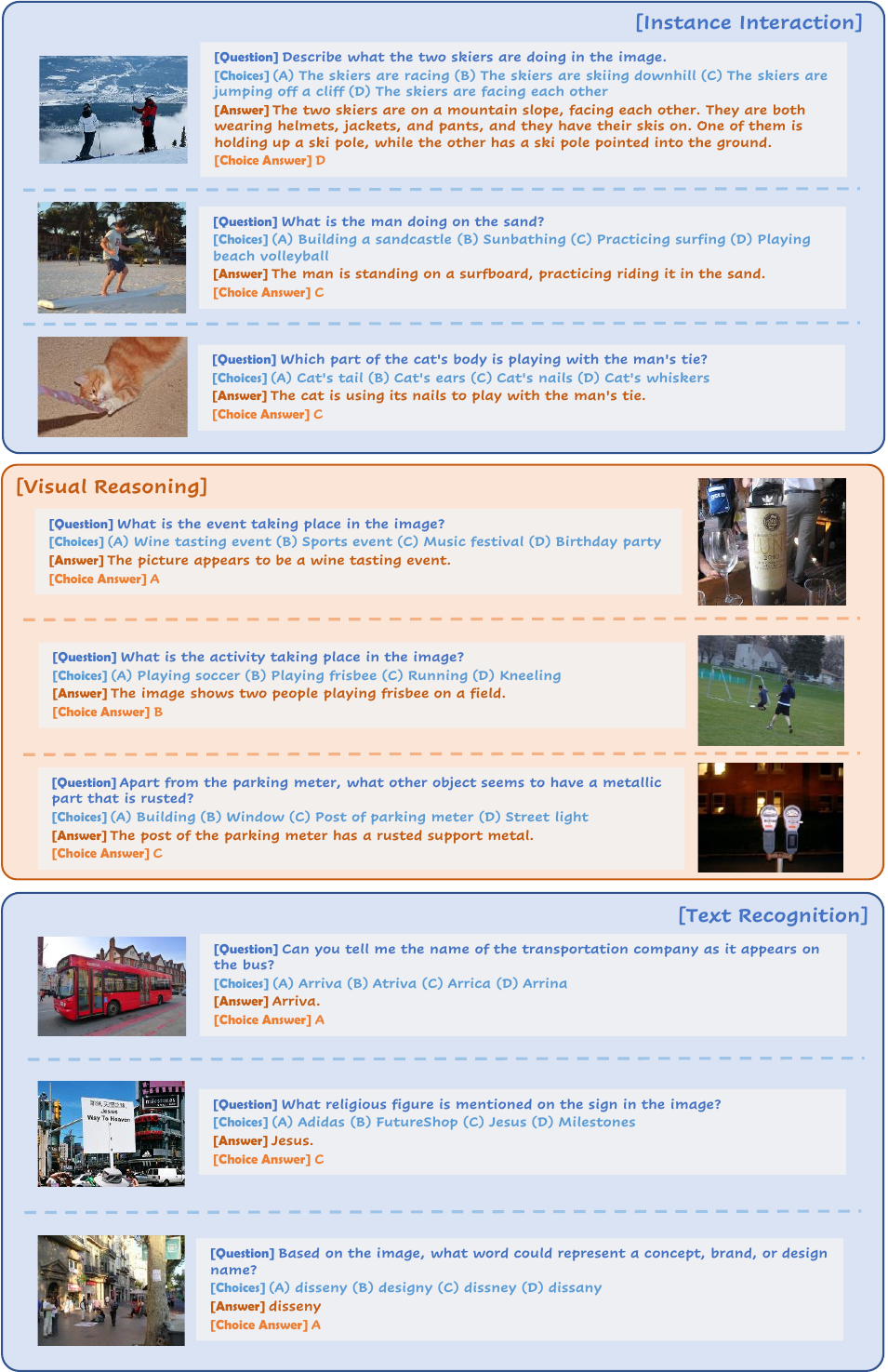}
    \caption{Examples of MLLM-DataEngine generated data in Instance Interaction, Visual Reasoning, and Text Recognition.}
    \label{fig:Showcase3}
\end{figure*}

High-quality and diverse examples generated by MLLM-DataEngine are demonstrated. Fig.~\ref{fig:Showcase1} are examples of Scene Understanding, instance Identity, and Instance Attribute. Fig.~\ref{fig:Showcase2} are examples of Instance Localization, Instance Counting, and Spatial Relation. Fig.~\ref{fig:Showcase3} are examples of Instance Interaction, Visual Reasoning, and Text Recognition. 



\bibliographystyle{IEEEbib}
\bibliography{icme2025references}

\newpage

